\documentclass{PoS}

\title{Zeros of the $W_L Z_L \rightarrow W_L Z_L$ amplitude:\\ With or without a light Higgs\thanks{This work has been supported in part by the Spanish Government 
and ERDF funds from the EU Commission [grants FPA2007-60323, 
FPA2011-23778, CSD2007-00042 (Consolider Project CPAN)].}}

\ShortTitle{Zeros of the $W_L Z_L \rightarrow W_L Z_L$ amplitude}

\author{A. Filipuzzi, \speaker{J. Portol\'es} and P. Ruiz-Femen\'{\i}a \\
        Departament de F\'{\i}sica Te\`orica, IFIC, CSIC --- Universitat de Val\`encia, \\ Apt. Correus 22085, E-46071 Val\`encia, Spain\\
        E-mail: \email{Alberto.Filipuzzi@ific.uv.es}, \email{Jorge.Portoles@ific.uv.es}, \email{Pedro.Ruiz@ific.uv.es} 
        }

\abstract{The existence of a new strong interacting sector around $E \sim 1 \, \mbox{TeV}$ is a common feature of Higgsless electroweak theories but also
of theories with a light Higgs, for instance, when this is not elementary. In those schemes, this new interaction could be at the origin of an extended
spectra with, in particular, spin-1 resonances that could be hinted in elastic gauge boson scattering.
Information on those resonances, if they exist, must be contained in the low-energy couplings of the electroweak chiral effective theory. Using the
facts that: i) the scattering of longitudinal gauge bosons, $W_L, Z_L$, can be well described in the high-energy region ($E \gg M_W$) by the
scattering of the corresponding Goldstone bosons (equivalence theorem) and that ii) the zeros of the scattering amplitude carry the information on the
heavier spectrum that has been integrated out; we employ the ${\cal O}(p^4)$ electroweak chiral Lagrangian, with or without a light Higgs state to identify
the parameter space region of the low-energy couplings where vector resonances may arise. }

\FullConference{The 7th International Workshop on Chiral Dynamics,\\
		August 6 -10, 2012\\
		Jefferson Lab, Newport News, Virginia, USA}

\begin{document}

\section{Introduction}
The breaking of the electroweak symmetry of the Standard Model (SM) is a major topic of research in particle physics as it encodes key aspects for our comprehension of 
the Universe, like the origin of mass or all the flavour physics. The hypothesis of a Higgs sector responsible for the spontaneous breaking of the symmetry has been taken along 
since almost the inception of the SM. However when LEP closed in 2000 the Higgs particle was still missing and we started to consider more plausible
that a Higgs sector could be absent, and the consequences this could bring into our understanding of particle physics were explored further.
\par 
Faced with the possibility of no Higgs, theoreticians have envisaged a Higgsless world where the spontaneous breaking of the electroweak symmetry would originate 
through other instances. One of the ideas considered the existence of a new strong interacting sector around $E \sim 1 \, \mbox{TeV}$ \cite{Veltman:1976rt,Lee:1977yc}. 
This new interaction would produce the breaking of the symmetry and, in the way, a complex spectra with resonance states, analogously to the low-energy Quantum Chromodynamics (QCD) case. 
The symmetry breaking sector of the Standard Model without a Higgs becomes a non-linear sigma model with $SU(2)_L \otimes SU(2)_R / SU(2)_V$ symmetry
where the $SU(2)_L \otimes U(1)_Y$ gauge symmetry is properly embedded. Interestingly enough the Lagrangian that describes it is the one of
two-flavour Chiral Perturbation Theory (ChPT) \cite{Weinberg:1978kz} with pions substituted by the Goldstone bosons
that provide masses to the gauge bosons. As in any effective field theory the low-energy coupling constants (LECs) of ChPT carry the information of the heavier
spectra that has been left out in the procedure of constructing the low-energy theory, as it has been proven at  ${\cal O}(p^4)$ in ChPT \cite{Ecker:1988te}. 
\par
The ATLAS and CMS experiments at the LHC have recently unveiled the existence of a boson with $M_H \sim 126 \, \mbox{GeV}$ \cite{:2012gk} 
that resembles very much the Higgs of the SM. Its properties and nature will be investigated in the next years.
There is also
the possibility that the new boson triggers the Higgs mechanism but is not an elementary particle. It could be a composite or a collective mode lying in an extended symmetry and in this case, again,
we would expect that this larger symmetry has new spectra in its linear representations.  
The above-mentioned setting, the non-linear sigma model, can also be modified in order to include a light Higgs boson that is a singlet under the $SU(2)_V$ symmetry above, 
called custodial \cite{Giudice:2007fh}. 
\par
In Ref.~\cite{Filipuzzi:2012bv} we proposed a procedure to explore the occurrence of spin-1 resonances in the $E \sim 1 \, \mbox{TeV}$ region 
based on the information about the spin-J resonances ($J \ge 1$) provided by the zeros of the scattering amplitude.
This method goes back to the study of the zeros in $\pi \pi \rightarrow \pi \pi$~\cite{Pennington:1973dz}.
In Ref.~\cite{Pennington:1994kc} it was shown in the framework of ChPT that the zeros of the isospin $I=1$ $\pi \pi \rightarrow \pi \pi$ amplitude
predict the mass of the $\rho(770)$ resonance when the chiral LECs are saturated by the
resonance contributions. This shows that, though the ChPT amplitude is only valid for $p^2 \ll M_{\rho}^2$, the extrapolation provided by 
its zeros is to be trusted up to $E \sim M_{\rho}$.
\par
This method can also be applied to the electroweak sector. On one side the elastic scattering amplitude of the longitudinal 
components of the gauge bosons is given, at $E \gg M_W$, by the amplitude of the
elastic scattering of the Goldstone bosons associated to the spontaneous electroweak symmetry breaking. This is known as the {\em equivalence theorem}
\cite{Lee:1977yc,Cornwall:1974km}.
This allows us to trade the dynamics of the longitudinally polarized gauge bosons by the one of the corresponding
Goldstone modes. The second ingredient is the fact that the interactions among Goldstone bosons is described,
at least at leading order, by the two-flavour ChPT Lagrangian where now the multiplet of
pions is substituted by the Goldstone fields that provide masses to the gauge bosons. The obvious difference is the relevant
scale that rules the perturbative expansion of the amplitude \cite{Longhitano:1980iz}. Indeed the perturbative
scale is now driven by $v \sim (\sqrt{2} \, G_F)^{-1/2} \simeq 246 \, \mbox{GeV}$ with $G_F$ the Fermi constant. Accordingly the effective
theory is  valid for $p^2 \ll (4 \pi v)^2 \sim \left( 3 \, \mbox{TeV} \right)^2$.  Taking into account the equivalence theorem,
the perturbative expansion and the electroweak chiral effective theory (EChET) our working region is
determined by $M_W \ll E \ll 4 \pi v$.

\section{The zeros of the scattering amplitude}
\label{sec:2}

As the ChPT amplitudes provide a perturbative expansion in momenta
it is clear that the resonances cannot be found as poles of amplitudes obtained in this approach. However a link between chiral dynamics and resonance
contributions can be provided employing some ad-hoc resummation techniques like Pad\'e approximants, the inverse amplitude method or the $N/D$
construction. We will propose an alternative procedure based on the zeros of the scattering
amplitude~\cite{Pennington:1973dz,Pennington:1994kc} as given by ChPT at ${\cal O}(p^4)$.
\par
Consider the amplitude $F(s,z)$ for $\pi^-(p_1) \pi^0(p_2) \rightarrow \pi^- \pi^0$ in the s-channel with $s=(p_1+p_2)^2$ and 
$z \equiv  \cos \theta = 1+2t/(s-4M_{\pi}^2)$. This amplitude has no $I=0$ component, and we know that the isovector P-wave is large
whereas the $I=2$ (exotic) S-wave is small. The P-wave is dominated by the $\rho(770)$ resonance and therefore around this energy region we can write the partial-wave expansion
of the amplitude as:
\begin{equation} \label{eq:pwfst}
 F(s,z) = 16\pi f_0^2(s) + \frac{48\pi}{\sigma}  \frac{M_{\rho} \Gamma_{\rho}(s)}{M_{\rho}^2-s-i M_{\rho} \Gamma_{\rho}(s)} \, z
+\,\dots,
\end{equation}
where $\sigma = \sqrt{1-4M_{\pi}^2/s}$  and $f_{\ell}^I(s)$ is the partial-wave with isospin $I$ and angular momentum $\ell$. 
\par
The dots in Eq.~(\ref{eq:pwfst}) amount to numerically suppressed higher partial waves.
Taking into account the small size of the S-wave component,
the angular distribution associated to $F(s,z)$ would have a marked dip at $z = 0$, where also $F(s,z)\simeq 0$. This reflects the spin-1
nature of the $\rho(770)$. Due to the properties of the Legendre polynomials these dips in the angular distribution (or zeros of the amplitude)
will appear for $\ell > 0$ and their number in
the physical region,  $z \in [-1,1]$, will be given by the angular momentum of the partial-wave. These zeros can be considered
as dynamical features which give the spin to the resonance. This observation provides a possible path to analyze the spectrum of $J \geq 1$ resonances
integrated out and hidden in the couplings of the effective field theory. 
\par
Being analytical functions of more than one variable the zeros of the amplitude are not isolated but continuous, defining a one-dimensional manifold for real $s$ and complex $t$.
Then the solution of $F(s,z_0) = 0$ for physical values of the $s$ variable is defined by
$z = z_0(s)$. We define the {\em zero contour} as the
real part of the zeros. This contour continues smoothly from one region to another in the Mandelstam plane.
Using Eq.~(\ref{eq:pwfst}) it can be seen that 
$\left| \mbox{Re}\,z(M_{\rho}^2) \right| \leq \frac{1}{3}$. Due to the exotic character of the
S-wave $I=2$ background and to the absence of the S-wave isoscalar channel,
we know that indeed $\left| \mbox{Re}\,z(M_{\rho}^2) \right| \ll\frac{1}{3}$.
\par
Hence, for a generic amplitude where the P-wave contribution dominates and is saturated by a vector resonance, the resonance mass $M_R$
should be found as the solution of:
\begin{equation} \label{eq:master}
\mbox{Re}\,z_0(M_R^2) \simeq 0 \, ,
\end{equation}
where $z_0(s)$ is the zero contour obtained from that amplitude. This procedure describes very accurately the $\rho(770)$ resonance 
starting with the elastic amplitude of $\pi \pi$ scattering  given by ${\cal O}(p^4)$ ChPT \cite{Filipuzzi:2012bv,Pennington:1994kc}.

\section{The Electroweak Chiral Lagrangian}
\label{sec:3}
In the absence of a Higgs, a strong interacting sector responsible for providing masses to the electroweak gauge bosons is described by Goldstone
bosons $\pi^a, a=1,2,3$, associated to the $SU(2)_L \otimes U(1)_Y \longrightarrow U(1)_{\mbox{em}}$ spontaneous symmetry breaking,
which become the longitudinal components of the electroweak gauge bosons.
The corresponding EChET Lagrangian is then described by the non-linear sigma model based on the coset
$SU(2)_L \otimes SU(2)_R / SU(2)_{L+R}$ where $SU(2)_L \otimes U(1)_Y$ is gauged. $SU(2)_{L+R}$ is
the custodial symmetry that is usually enforced in order
to keep the relation $M_W = M_Z \cos \theta_W$ and the smallness of the $T$ oblique parameter.
A convenient parameterization of the Goldstone fields is given by
$
 U(x) = \exp \left( \frac{i}{v} \, \pi^a \tau^a \right) 
$,
with $\tau^a$ the Pauli matrices. This transforms as $L U R^{\dagger}$, with $L \in SU(2)_L$ and $R \in U(1)_Y$, under the gauge group.
Up to dimension four operators, the most general $SU(2)_L \otimes U(1)_Y$ gauge and CP invariant
Lagrangian which implements the global symmetry breaking
$SU(2)_L \otimes SU(2)_R$ into $SU(2)_{L+R}$ is given in Ref.~\cite{Longhitano:1980iz}.
\par
A light Higgs boson, singlet of the $SU(2)_{L+R}$ custodial symmetry, can be accomodated in this non-linear
sigma model by multiplying the operators 
by arbitrary polynomials of the Higgs field $f_i(H)$. We keep here only those operators needed for our task:
\begin{equation} \label{eq:echet}
 {\cal L}_{\mbox{\footnotesize EChET}} = \frac{v^2}{4} \langle \left( D_{\mu} U \right)^{\dagger} D^{\mu} U \rangle f_0(H)  +
 a_4 \, \langle V_{\mu} V_{\nu} \rangle^2 f_4(H) \, + a_5 \, \langle V_{\mu} V^{\mu} \rangle^2 f_5(H) \, , 
\end{equation}
where $V_{\mu} = \left( D_{\mu} U \right) U^{\dagger}$ and $f_i(0) = 1$. In order to study the scattering of longitudinally
polarized gauge bosons, when the Higgs of the SM is also included, we can set $f_4(H)=f_5(H)=1$ while $f_0(H) = \left( 1 + 2 H/v\right)$.
The covariant derivative takes the form
$
 D_{\mu} U = \partial_{\mu} U + \frac{i}{2} \, g \, \tau^k \, W_{\mu}^k U - \frac{i}{2}\,  g'\,  \tau^3 \, U B_{\mu} 
$.
${\cal L}_{\mbox{\footnotesize EChET}}$ involves a
perturbative derivative expansion driven this time by the scale
$\Lambda_{\mbox{\footnotesize EW}} = 4 \pi v \simeq 3 \, \mbox{TeV}$, {\it i.e.} an expansion in
powers of $(p^2,M_V^2)/\Lambda_{\mbox{\footnotesize EW}}^2$.
The low-energy couplings $a_4$ and $a_5$
encode the information of the heavier spectrum that has been integrated out in order to
get ${\cal L}_{\mbox{\footnotesize EChET}}$. 
\par
Here we are interested in the scattering of vector bosons with longitudinal polarization
because is the one linked, through the Higgs mechanism, with the Goldstone bosons of the electroweak symmetry breaking sector.
The exact relation is provided by the {\em equivalence theorem}~\cite{Lee:1977yc,Cornwall:1974km}:
\begin{equation} \label{eq:et2}
 A \left( V^a_L V^b_L \rightarrow V^c_L V^d_L \right)
= 
A^{(4)} \left( \pi^a \pi^b \rightarrow \pi^c \pi^d \right)
+ {\cal O}\left( \frac{M_V}{E} \right)
+ \, {\cal O}(g,g^\prime) + {\cal O} \left( \frac{E^5}{\Lambda_{\mbox{\footnotesize EW}}^5} \right)\, ,
\end{equation}
where $A^{(4)}$ is the amplitude of Goldstone boson scattering at lowest order in the electroweak
couplings ($g$ and $g^\prime)$ as obtained from the effective Lagrangian
${\cal L}_{\mbox{\footnotesize EChET}}$.
Let us remark that at ${\cal O}(g^0, g'^0)$ the masses of the gauge bosons
vanish and the equivalence theorem indicates that the Goldstone boson
scattering amplitude has to be calculated in the zero mass limit.
\par
Notice that Eq.~(\ref{eq:et2}) is valid in the energy range given by
$M_V \ll E \ll \Lambda_{\mbox{\footnotesize EW}}$.
Since the EChET framework is analogous to ChPT, and the latter works reasonably well
up to $500$~MeV $\simeq 2\pi F_\pi$ , we can assume that our effective formalism
could be valid at least up to $2\pi v\simeq 1.5$~TeV.

\section{Analysis of the zeros of the $W_L Z_L \rightarrow W_L Z_L$ amplitude}
\label{sec:4}

The equivalence theorem can be used to relate, at leading order, the amplitude of $W_L Z_L \rightarrow W_L Z_L$ with
the one of the corresponding Goldstone bosons. The physical system provided by the Lagrangian
in Eq.~(\ref{eq:echet}) withouth Higgs is,
but for the change of scale ($F_{\pi} \rightarrow v$), the same than the one of the ChPT. Therefore one would expect a similar dynamics
if the P-wave contribution is saturated by a vector resonance. Consequently
we could apply the same procedure and study the occurrence of vector resonances in the scattering $W_L Z_L \rightarrow W_L Z_L$
through the analysis
of the zero contours of the EChET amplitude. Zero contours cross the resonance location
close to where the Legendre polynomial vanishes, which for vector resonances amounts to the condition (\ref{eq:master}).
\par
The scale-independent $\bar{a}_i$ couplings are related to
their renormalized counterparts in the $\overline{\mbox{MS}}$ scheme as 
$a_i^r(\mu) = \frac{1}{N_i} \left( \overline{a}_i -1 + \ln \frac{M_W^2}{\mu^2} \right)$ for $i=4,5$ with 
$N_4 = 192 \pi^2$ and $N_5 = 2 N_4$.
The natural order of magnitude of the couplings is
$\bar{a}_{4,5}\sim {\cal O}(1)$, so we can expect that $a_i^r\sim {\cal O}(10^{-3})$.
\par
We look for the zero contours of the amplitude,
$A^{(4)}(s,z_0)=0$, and identify the vector resonances with solutions
of $\mbox{Re}\,z_0(M_R^2)=0$.
At the resonance location, $s=M_R^2$, the latter equation relates the
size of the imaginary part of the zero  with the ratio between the S- and the P-wave contributions:
\begin{equation} \label{eq:imb}
\left| z_0(M_R^2) \right|  = \left| \mbox{Im} \,z_0(M_R^2) \right| = \left| \frac{f_0^2(M_R^2)}{3 \, f_1^1(M_R^2)} \right| < \lambda \, .
\end{equation}
The bound $\lambda$ then defines the range of applicability of our method: zeros of the amplitude with imaginary part smaller than $\lambda$ can be
 considered positive results in the search for vector resonances. 
 For values of $\lambda$ larger than $1/2$ we cannot consider the S-wave to be significantly smaller than
the P-wave, and one of the hypothesis of our method is not fulfilled~\cite{Filipuzzi:2012bv}.
\begin{figure}[h!p]
\begin{center}
\includegraphics[width=11cm]{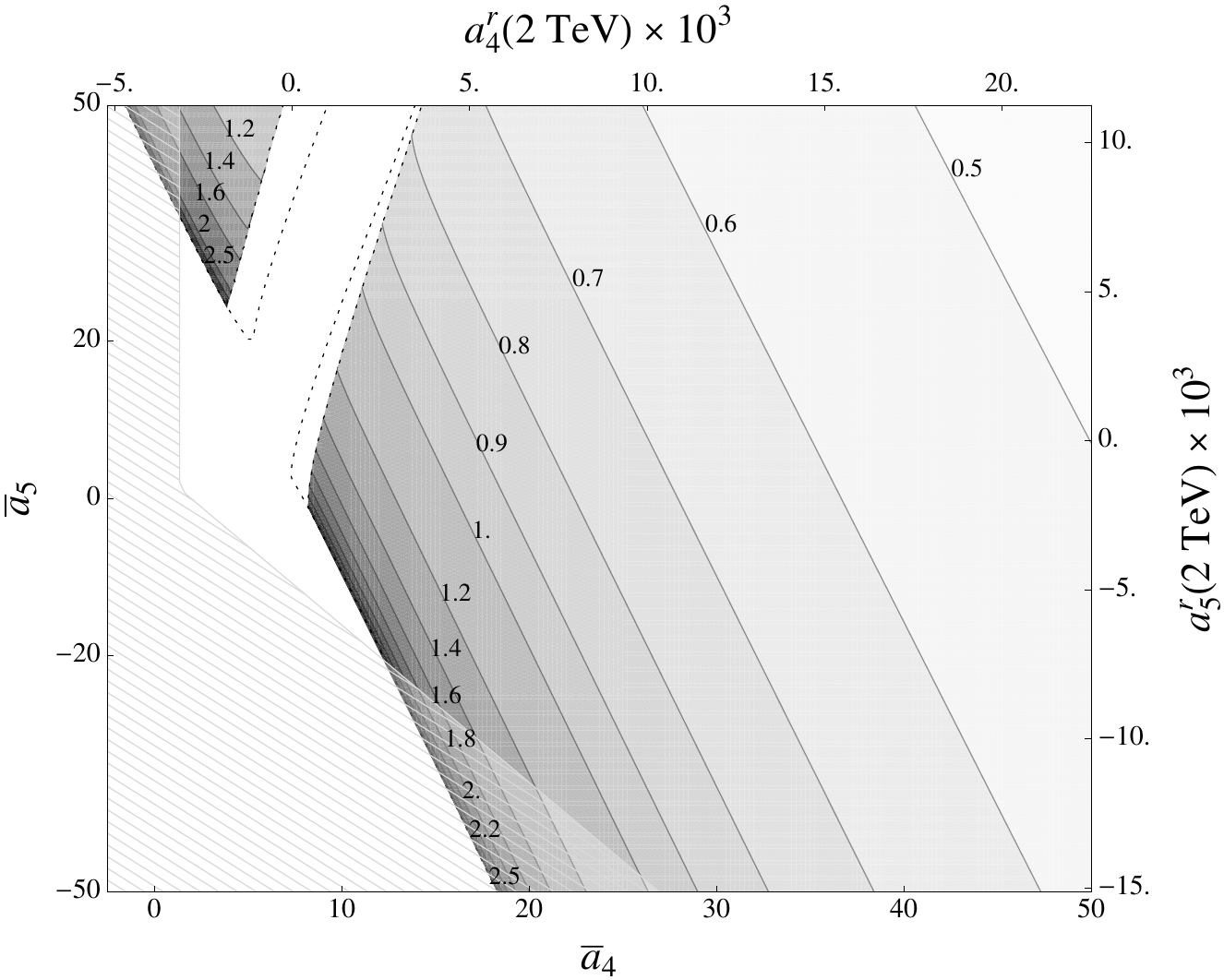}
\caption{\label{fig:4} Resonance masses as a function of the low-energy couplings $\bar{a}_4$ and $\bar{a}_5$ in a world without a light Higgs. 
The scales in terms of
the renormalized couplings $a_4^r(\mu)$ and $a_5^r(\mu)$ at $\mu=2$~TeV are also drawn.
The shaded areas show where resonances defined by the conditions~(\protect\ref{eq:master}) and~(\protect\ref{eq:imb})
with $\lambda=1/3$ are found in the $(\bar{a}_4,\bar{a}_5)$-plane. The contour lines drawn correspond to 
pairs of $(\bar{a}_4,\bar{a}_5)$ which yield the same resonance mass. The hatched region corresponds to values of
$\bar{a}_4$ and $\bar{a}_5$ forbidden by positivity conditions on the $\pi\pi$ scattering amplitudes.
The outermost dashed lines mark the boundary of the resonance region corresponding to $\lambda=1/2$.}
%
\includegraphics[width=11cm]{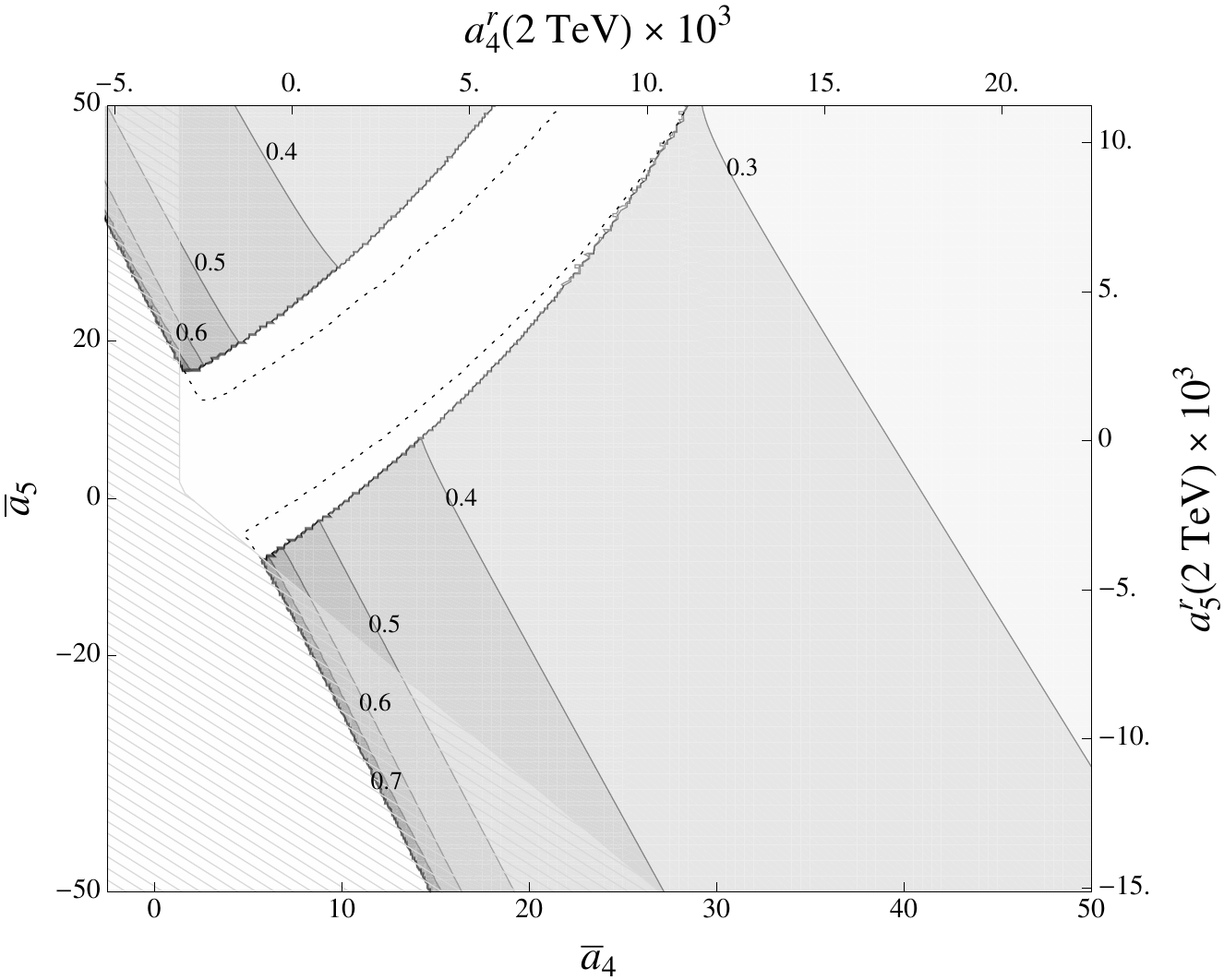}
\caption{\label{fig:5} Resonance masses as a function of the low-energy couplings $\bar{a}_4$ and $\bar{a}_5$ when the Higgs of the SM 
with a mass of $M_H = 126 \, \mbox{GeV}$ is included in the analysis. Further explanations are given in Figure~\protect\ref{fig:4}.}
\end{center}
\end{figure}
\par
We now consider two settings~: the Higgsless case shown in Figure~\ref{fig:4} \cite{Filipuzzi:2012bv}, and the SM with a light Higgs,
with mass $M_H = 126 \, \mbox{GeV}$
shown in Figure~\ref{fig:5}. Though the validity
of the approach cannot be trusted beyond $E \simeq 2$~TeV, we have displayed in the plot resonances found with masses up to $2.5$ TeV.
On the other side, the use of the equivalence theorem indicates that very low masses, let us say $M_R \lesssim 0.5 \, \mbox{TeV}$ could
also be at odds with our procedure. 
In order to study de $\lambda$ dependence we consider two cases, $\lambda=1/2$ and $\lambda=1/3$.
The hatched region in the left and lower parts of the plot corresponds to values of
$\bar{a}_4$ and $\bar{a}_5$ forbidden by positivity conditions on the $\pi\pi$ scattering amplitudes \cite{Pennington:1994kc}.
\par
Let us comment the most relevant features of our results: \vspace*{0.1cm} \\
\underline{\bf{ 1/} \emph{Higgsless case}}. As can be seen in Figure~\ref{fig:4}, no vector resonances are found for $\bar{a}_4\lesssim 8$ and $\bar{a}_5\lesssim 25$.
This would exclude to a large extent Higgsless models with vector resonances which saturate the low-energy couplings to the
expected natural order of magnitude ($\bar{a}_{4,5}\sim 1$). Masses above $1.8$~TeV are confined to a thin slice in the lower-left and upper-left parts of the shaded regions
and are mostly excluded by the positivity constraints. Conversely, light resonances ($M_R \lesssim 0.8$~TeV) require values of either $\bar{a}_4$ or $\bar{a}_5$ larger than 20. 
The validity of the EChET Lagrangian for such large values of the LECs is nevertheless questionable. \vspace*{0.1cm} \\
\underline{\bf{2/} \emph{SM Higgs}}. Including the Higgs contribution in the Goldstone boson scattering amplitude has a large effect on 
the location of the zeros.
From Figure~\ref{fig:5} we see that vector resonances are only found for $\bar{a_i}$ outside the natural order of
magnitude region, and that all possible
vector resonances have now a light mass,  $M_R \lesssim 0.7 \, \mbox{TeV}$, thus lying close to the energy region where the method of zeros 
is not valid. Hence, apart from the narrow slice of  $\bar{a_i}$ values where  $M_R \gtrsim 0.5 \, \mbox{TeV}$, 
our method excludes 
the existence of vector resonances with masses $ 0.5 \, \mbox{TeV} \lesssim M_R \lesssim 1.5 \, \mbox{TeV}$,
in agreement with 
the results of Ref.~\cite{Pich:2012dv}.
\par
In both cases the dependence on the $\lambda$-cut, though noticeable, is small.
The LHC sensitivity to explore the values of the coefficients $a_4$ and $a_5$ has been investigated in
Ref.~\cite{Eboli:2006wa}. 
Using ATLAS and CMS data, the most recent bounds on the mass of new charged vector resonances have also been obtained
\cite{Chatrchyan:2012meb,Chatrchyan:2012kk}.
Depending on the model, it is possible to exclude resonances with a mass as heavy as 1.14 and 2.5 TeV, 
in the $WZ$ and leptonic case respectively.  
\par
A systematic study of resonance masses in the parametric space spanned by $a_4^r(\mu)$ and $a_5^r(\mu)$ using the Inverse Amplitude Method  has 
also been performed both in the Higgsless case~\cite{Dobado:1999xb} and including a light SM Higgs \cite{Espriu:2012ih}. Their results, compared with
both our Figures~\ref{fig:4} and \ref{fig:5}, look rather different.

\end{document}